\documentclass[aps,prl,preprint,groupedaddress,showpacs]{revtex4}
\usepackage[dvips]{graphicx}
\usepackage{longtable}
\usepackage{bm}

\begin{document}

\title{Consistent $\alpha$-cluster description of 
the $^{12}\mathrm{C} (0^{+}_2)$ resonance} 

\author{S.~I.~Fedotov} 
\author{O.~I.~Kartavtsev} 
\author{A.~V.~Malykh} 
\affiliation{Joint Institute for Nuclear Research, 141980, Dubna, Russia} 

\date{\today}

\begin{abstract} 

The near-threshold $^{12}\mathrm{C}(0^{+}_2)$ resonance provides unique 
possibility for fast helium burning in stars, as predicted by Hoyle to explain 
the observed abundance of elements in the Universe. 
Properties of this resonance are calculated within the framework of 
the $\alpha$-cluster model whose two-body and three-body effective potentials 
are tuned to describe the $\alpha - \alpha$ scattering data, 
the energies of the $0^+_1$ and $0^+_2$ states, and the $0^+_1$-state 
root-mean-square radius. 
The extremely small width of the $0^{+}_2$ state, the $0_2^+ \to 0_1^+ $ 
monopole transition matrix element, and transition radius are found in 
remarkable agreement with the experimental data. 
The $0^{+}_2$-state structure is described as a system of three 
$\alpha$-particles oscillating between the ground-state-like configuration and 
the elongated chain configuration whose probability exceeds 0.9. 
\end{abstract} 

\maketitle 


The extremely narrow $^{12}\mathrm{C}(0^{+}_2)$ resonance just above 
the 3$\alpha$ threshold was predicted by Hoyle~\cite{Hoyle54} to explain 
the observed abundance of elements in the Universe. 
Following the prediction, the near threshold 3$\alpha$ 
($^{12}\mathrm{C}(0^{+}_2)$) and 2$\alpha$ ($^8\mathrm{Be}$) resonances 
are crucial for sufficiently fast carbon production in stellar 
nucleosynthesis which goes through the reactions $3\alpha \to {^8\mathrm{Be}} 
+ \alpha \to {^{12}\mathrm{C}}(0_2^+) \to {^{12}\mathrm{C}} + \gamma$. 
The experimental confirmation came soon~\cite{Cook57} and properties of this 
state have been fairly well studied experimentally during the following 50 
years. 
In particular, the extremely small width $\Gamma$ and the $0_2^+ \to 0_1^+ $ 
transition density (the monopole transition matrix element $M_{12}$ and 
the transition radius $R_{12}$) were determined. 
Besides the resonance triple-$\alpha$ reaction, a consistent treatment of 
the three-body near-threshold dynamics is ultimately necessary to study 
the non-resonance reaction, which is of importance at low temperature 
and high helium density, as occurs in accretion on white dwarfs and neutron 
stars~\cite{Cameron59}. 
As the experimental data at these conditions are lacking, the reliable 
theoretical results are especially desired; nevertheless, the calculated 
non-resonance reaction 
rates~\cite{Nomoto85,Fushiki87,Schramm92,Ogata09} differ by many orders of 
magnitude. 
There is vast literature on the $^{12}\mathrm{C}$ states above the 3$\alpha$ 
threshold, and it is possible to notice just some of the recent papers, e.~g., 
with concern of astrophysical aspects~\cite{Oberhummer00,Fynbo05,Ogata09}, 
distribution of decaying particles~\cite{Alvarez-Rodriguez08,Kirsebom10}, 
dilute-gas variational calculations~\cite{Funaki06}, and different approaches 
to precise description of the $^{12}\mathrm{C}$ 
properties~\cite{Fedotov04,Fedotov05,Filikhin05,Chernykh07,Descouvemont10}. 
Besides much interest in the $^{12}\mathrm{C}(0^{+}_2)$ state owing to its 
role in the astrophysical applications, the structure of this state has been 
widely discussed for a long time, in particular, a linear chain configuration 
was suggested in~\cite{Brink70} and a triangle configuration was considered 
in~\cite{Bijker00}. 

The following reasons impede the theoretical description of 
the $^{12}\mathrm{C}(0^{+}_2)$ resonance and, more generally, of 
the 3$\alpha$ low-energy processes. 
Application of the twelve-nucleon calculations is hardly feasible for 
the unstable near-threshold $0^+_2$ state; the extremely difficult task is 
to describe the $\alpha$-cluster structure and complicated three-body dynamics 
of $\alpha$-particles. 
On the other hand, although the $\alpha$-cluster model is a natural 
alternative and a number of works have elaborated in this direction, it 
remains to prove that the model would provide the desired accuracy. 
One of the crucial difficulties stems from the three-body decay of the $0^+_2$ 
resonance, which requires the construction of the asymptotic wave function for 
three charged particles. 

The present Letter is aimed at reliable calculation of 
the $^{12}\mathrm{C}(0^+_2)$ resonance within the framework of 
the $\alpha$-cluster model with particular intention to obtain its extremely 
small width $\Gamma$ and the experimentally observable monopole transition 
matrix element $M_{12}$ and transition radius $R_{12}$. 
Moreover, the reliable $\alpha$-cluster calculation will be used to elucidate 
the $0^+_2$-state structure. 
The $\alpha$-cluster model is applied also to calculate the $0^+_1$ state 
despite that its cluster structure is not justified. 
First, this calculation provides a consistent treatment of 
the $0^{+}_2$--$0^{+}_1$ transition as the wave functions of both states 
should be calculated within the same approach. 
Next, it is of importance for a correct description of those effects which 
originate from the $\alpha$-particles structure. 
An additional goal of the $^{12}\mathrm{C}(0^+_2)$ calculation is to construct 
the effective potentials of the $\alpha$-cluster model which is known to be 
successful in different nuclear problems. 
The focus of the present calculation is the precise solution of the three-body 
problem which allows one to show an ultimate ability of the $\alpha$-cluster 
model in description of the fine $^{12}\mathrm{C}$ characteristics. 

For a precise description of the $^{12}\mathrm{C}$ nucleus, firstly, 
the effective three-body interaction is introduced to take into account 
the finite size and nucleonic structure of the $\alpha$-particles including 
the effects of wave-function antisymmetrization under nucleons. 
Next, the assumption on the ``frozen'' nucleon distribution in 
the $\alpha$-particles is used to calculate the experimentally observable 
nucleonic structural characteristics, e.~g., the root-mean-square (r.~m.~s.) 
radii and the transition radius $R_{12}$. 
This extended $\alpha$-cluster model leads generally to a satisfactory 
description of the $^{12}\mathrm{C}$ nucleus 
\cite{Fedotov04,Fedotov05,Filikhin05,Alvarez-Rodriguez07,Alvarez-Rodriguez08}. 
In particular, the calculated $\Gamma$ and $M_{12}$~\cite{Fedotov04,Fedotov05} 
are in reasonable agreement with experimental data (up to a factor two) even 
for the simplified local $\alpha$ - $\alpha$ interactions. 

The effective potentials of the $\alpha$-cluster model are constructed under 
the following natural requirements. 
The two-body potential is chosen to reproduce the low-energy 
elastic-scattering phase shifts in the lowest even-parity waves and to fix at 
the experimental values both energy and width of the narrow $\alpha$-$\alpha$ 
resonance (the $^8\mathrm{Be}$ ground state). 
The latter condition is essential for a correct description of 
the multi-dimensional potential barrier whose penetrability determines 
the extremely small width $\Gamma$ of the $^{12}\mathrm{C}(0_2^+)$ state. 
Furthermore, for any reasonable calculation of the width $\Gamma$, 
the $0^+_2$-state energy should be precisely fixed to the experimental value 
by a proper choice of the effective three-body potential. 
As the ground-state wave function is needed to treat the $0^{+}_2$--$0^{+}_1$ 
transition, the effective three-body potentials should provide also 
the experimental values of the energy and r.~m.~s. radius of the $0_1^+$ 
state. 


The Schr\"odinger equation reduces to the system of coupled hyperradial 
equations (HRE) for the channel functions $f_n(\rho)$ by using the expansion 
of the three-body wave function in a set of eigenfunctions 
$\Phi_n(\alpha, \theta; \rho)$ on a hypersphere at fixed $\rho$, 
\begin{equation}
\Psi = \rho^{-5/2}\displaystyle\sum_{n = 1}^{\infty}
f_n(\rho)\Phi_n(\alpha, \theta; \rho) \, . 
\label{expansion}
\end{equation}
Here the hyperspherical variables $0 \leq \rho < \infty$, 
$0 \leq \alpha_i \leq \pi$, and $0 \leq \theta_i \leq \pi$ are defined by 
$x_i = \rho \cos(\alpha_i/2)$, $y_i = \rho \sin(\alpha_i/2)$, and 
$\cos\theta_i = ({\mathbf x}_i \cdot {\mathbf y}_i)/(x_i y_i)$, the scaled 
Jacobi coordinates are ${\mathbf x}_i = {\mathbf r}_j - 
{\mathbf r}_k ,\ {\mathbf y}_i = (2{\mathbf r}_i - {{\mathbf r}_j - 
{\mathbf r}_k})/\sqrt{3}$ and ${\mathbf r}_i$ is the position of the $i$th 
$\alpha$-particle. 
Both the total wave function $\Psi$ and the eigenfunctions 
$\Phi_n(\alpha, \theta; \rho)$ are symmetrical under any permutation 
of the $\alpha$-particles. 
In more detail, the method of calculation and the numerical procedure are 
presented in~\cite{Fedotov04,Fedotov05}. 

The two-body interaction includes the Coulomb part, $4 e^2/x$, and 
the short-range part of the simple Ali-Bodmer form, 
\begin{equation}
\label{eff3}
V(x) = V_r e^{-x^2/\mu_r^{2} } + V_a e^{-x^2/\mu_a^{2} } \, , 
\end{equation} 
in the $s$, $d$, and $g$ partial waves. 
Following the above discussed requirements, the parameters of the two-body 
potentials are chosen to fit the low-energy $\alpha$ - $\alpha$ 
elastic-scattering phase shifts (up to $E_{\rm cm} = 12$~MeV) and to reproduce 
the experimental energy $E_{2\alpha} = 92.04 \pm 0.05$~keV and width 
$\gamma = 5.57 \pm 0.25 $~eV of the $^8\mathrm{Be}$ ground 
state~\cite{Wustenbecker92,Tilley04}. 
These conditions do not uniquely fix the two-body interactions and 
to elucidate its role two variants are considered for each of the $s$-, $d$-, 
and $g$-wave potentials whose parameters are listed in Table~\ref{2bpot}. 
\begin{table}[htb]
\begin{ruledtabular}
\begin{tabular}{c@{\hspace{10bp}}c@{\hspace{5bp}}c@{\hspace{8bp}}
c@{\hspace{5bp}}c@{\hspace{8bp}}c@{\hspace{5bp}}c} 
 & \multicolumn{2}{c}{$l = 0$} & \multicolumn{2}{c}{$l = 2$} & 
\multicolumn{2}{c}{$l = 4$} \\ 
 & s0 & s1 & d0 & d1 & g0 & g1 \\
\hline
$V_r $   & 234.914  & 295.160  & 152.9    & 240.0    & 10.0    & 36.0 \\
$\mu_r $ & 1.54     & 1.4213   & 1.4213   & 1.3      & 1.424   & 1.424   \\
$V_a$    & -109.766 & -99.1406 & -99.1406 & -99.1406 & -134.0  & -140.0 \\
$\mu_a $ & 2.0944   & 2.09455  & 2.09455  & 2.09455  & 2.09455 & 2.09455 
\end{tabular}
\end{ruledtabular}
\caption{Parameters $V_{r, a} $~(MeV) and $\mu_{r, a}$~(fm) of 
the $\alpha$-$\alpha$ potential~(\ref{eff3}) for two variants in each of 
the $l = 0, 2, 4$ partial waves. }
\label{2bpot}
\end{table}
Given the two-body potential, the corresponding three-body 
potential of the Woods-Saxon form, 
\begin{equation}
V_{3}(\rho) = V_0[1 + e^{(\rho - a)/b}]^{-1} \, , 
\label{ef5}
\end{equation} 
is chosen to fix the $0_1^+$-state energy, the $0_2^+$-state energy, and 
the $0_1^+$-state r.~m.~s. radius at the experimental values 
$E_1 = -7.2747$~MeV, $E_2 = 0.3795$~MeV~\cite{Ajzenberg90}, and 
$R_1 = 2.48 \pm 0.002$~\cite{Offermann91,Ruckstuhl84}. 
Five of eight possible combinations of the two-body potentials under 
consideration, which provide the desired $E_1$, $E_2$, and $R_1$, and fitted 
parameters $V_0$, $a$, and $b$ of the three-body potential~(\ref{ef5}) are 
presented in Table~\ref{tabprop}. 
\begin{table}[htb]
\begin{ruledtabular}
\begin{tabular}{c@{\hspace{7bp}}c@{\hspace{5bp}}c@{\hspace{5bp}}
c@{\hspace{6bp}}c@{\hspace{5bp}}c@{\hspace{5bp}}c@{\hspace{5bp}}c}
 & $V_0$ & $a$ & $b$ & $\Gamma$ & $R_2$ & $M_{12}$ & $R_{12}$ \\
\hline
s0 d0 g0 & -33.7737 & 2.87500 & 0.97565 & 8.29 & 3.555 & 5.270 & 4.844  \\
s0 d0 g1 & -50.5656 & 2.23438 & 1.05366 & 8.51 & 3.572 & 5.278 & 4.852  \\
s0 d1 g0 & -129.031 & 0.96875 & 1.11313 & 8.65 & 3.576 & 5.449 & 4.847  \\
s1 d0 g0 & -63.4126 & 1.90625 & 1.06614 & 7.92 & 3.574 & 5.335 & 4.836  \\
s1 d0 g1 & -109.009 & 1.15234 & 1.1075  & 8.71 & 3.590 & 5.298 & 4.843  
\end{tabular}
\end{ruledtabular}
\caption{The $0_2^+$ state width $\Gamma$~(eV), the r.~m.~s. radius 
$R_2$~(fm), the monopole transition matrix element $M_{12}$~(fm$^2$), and 
the transition radius $R_{12}$~(fm). 
Parameters of the two-body potential indicated according to 
Table~\protect\ref{2bpot} and parameters of the corresponding three-body 
potential~(\ref{ef5}) are $V_0$~(MeV), $a$~(fm), and $b$~(fm). } 
\label{tabprop}
\end{table}

A set of eigenfunctions $\Phi_n(\alpha, \theta; \rho)$ and all the terms 
in the HRE are obtained by using the variational method with a flexible basis 
of symmetrical (under any permutation of three particles) trial functions. 
For small $\rho \leq 20$~fm, sufficient precision is achieved on the basis 
containing 192 (corresponding to $K_{\mathrm{max}} = 90$) symmetrical 
hyperspherical harmonics (SHH). 
With increasing $\rho$, a number of SHH providing the desired accuracy 
enormously grows due to the $\alpha + \, ^8\mathrm{Be}$ cluster structure of 
the wave function. 
Therefore, the desired precision is obtained with the basis containing 
a set of 108 SHH (corresponding to $K_{\mathrm{max}} = 66$) and four 
additional  trial functions $\varphi_j = \sum_i \exp{(-\beta_j x_i^2)}$. 

\begin{figure}[htb]
\includegraphics[width=0.8\textwidth, angle=-90]{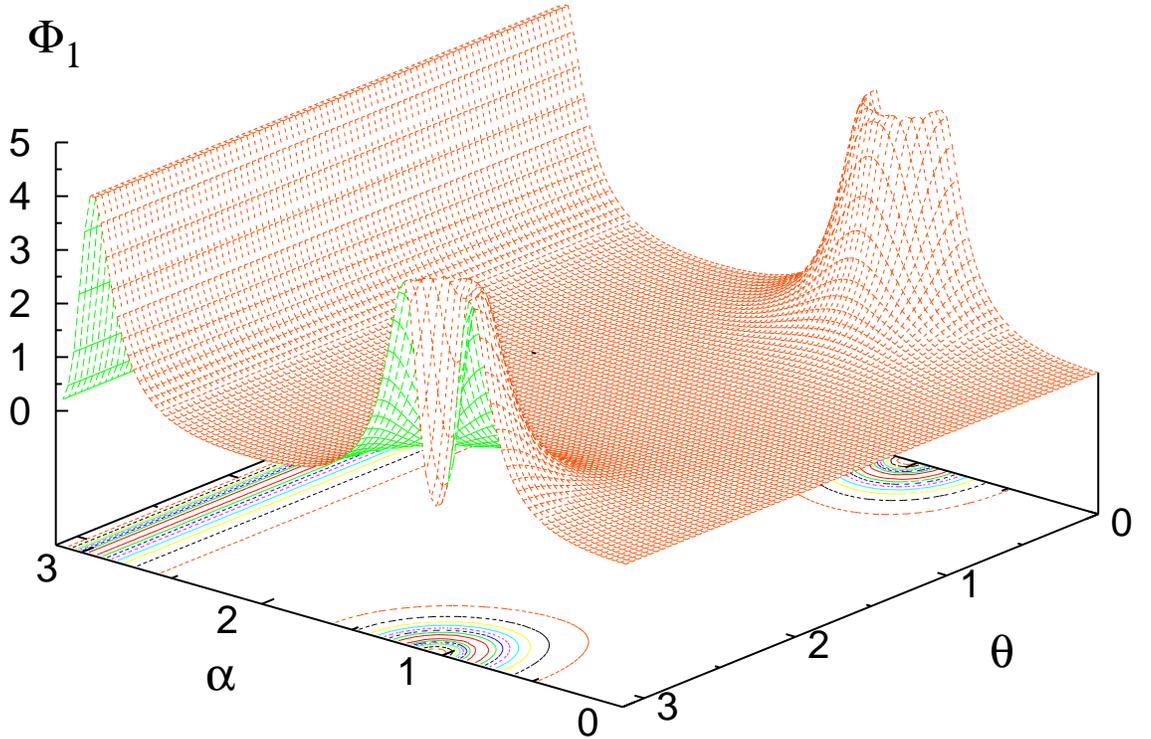}
\caption{The eigenfunction $\Phi_1(\alpha, \theta; \rho)$ at $\rho = 30$~fm.} 
\label{fig1}
\end{figure}

Whereas the simple zero asymptotic boundary conditions in the HRE are used in 
the calculation of the $0_1^+$ state, the complicated scattering problem for 
three charged particles should be solved to describe the $0_2^+$ resonance. 
Fortunately, the $\alpha + \, ^8\mathrm{Be}$ two-cluster structure of 
the wave function at large, though finite, distances is corroborated in 
the present calculations, which allows one to circumvent the tremendous 
calculation of the three-charged-particle wave function. 
The $\alpha + \, ^8\mathrm{Be}$ structure follows from the form of 
$\Phi_1(\alpha, \theta; \rho)$ which is close to the symmetrized combination 
$\sum_i \phi(x_i)$ of the $^8\mathrm{Be}$ wave function $\phi(x)$ in a wide 
region $20$~fm $\le \rho \le 70$~fm, where the overlap integral on 
a hypersphere $\langle \Phi_1 | \sum_i \phi(x_i) \rangle $ exceeds $0.977$. 
This form of $\Phi_1(\alpha, \theta; \rho)$ is demonstrated in Fig.~\ref{fig1} 
for $\rho = 30$~fm, where three peaks represent the functions $\phi(x_i)$. 
The two-cluster structure is approved also by the form of the first-channel 
effective potential in the HRE, which to good accuracy is a sum of 
the ${^{8}\mathrm{Be}}$ energy $E_{2\alpha}$ and the Coulomb interaction 
between $\alpha $ and ${^{8}\mathrm{Be}}$, $2\eta e^2/\rho $ in the interval 
$40$~fm $ < \rho < 70$~fm. 
Indeed, fit to this dependence gives $89.29$~keV $ < E_{2\alpha} < 90.4$~keV 
and $6.73$ $ < \eta < 6.76$ which are close to the expected values 
$E_{2\alpha} = 92.04 \pm 0.05$~keV and $\eta = 8/\sqrt{3} \approx 6.65$. 
Thus, the $0_2^+$-state properties are calculated by using the two-cluster 
($\alpha + \, ^8\mathrm{Be}$) asymptotic boundary condition for 
the first-channel function $f_1(\rho)$ at a large, though finite, hyperradius. 
More precisely, the asymptotic form is 
$f_1(\rho) \propto F_0(\eta/k r_b, k\rho) + \tan\delta(E) 
G_0(\eta/k r_b, k\rho)$, where $\delta(E)$ is the scattering phase shift, 
$F_0$ and $G_0$ are the Coulomb functions, $k^2 = 
(m_\alpha/\hbar^2)(E - E_{2\alpha})$, and $r_b = \hbar^2/(m_\alpha e^2)$. 
The $0_2^+$-state energy $E_2$ and width $\Gamma$ are obtained by fitting 
$\delta(E)$ to the Wigner energy dependence. 

The assumption on the fixed nucleon distribution in the $\alpha$-particle 
leads to the representation of the twelve-nucleon wave function as a product 
of the three-body wave function $\Psi $ and the internal wave functions of 
the three $\alpha$-particles. 
Using this form, one obtains from the usual 
definitions~\cite{Fedotov04,Fedotov05,Strehl68} that the monopole transition 
matrix element is expressed merely via the three-body wave functions, 
$M_{12} = \langle\Psi^{(1)}\left| \rho^2 \right|\Psi^{(2)} \rangle $, 
whereas the r.~m.~s. radii and the transition radius depend also on 
the $\alpha$-particle r.~m.~s. radius 
$R_{\alpha} = 1.681\pm 0.004$~fm~\cite{Sick08}, viz., 
$R_i^{2} = \frac{1}{6} \langle \Psi^{(i)} \left| \rho^2 \right| \Psi^{(i)} 
\rangle + R^2_{\alpha} $ 
and
$R^2_{12} =  \frac{\langle \Psi^{(1)}\left| \rho^4 
\left( 3 - \sin^2\alpha \sin^2\theta \right) \right|\Psi^{(2)} \rangle }
{12 M_{12}} + \frac{10}{3} R^2_{\alpha} $. 
In calculation of $M_{12}$, $R_{2}$, and $R_{12}$ the ultra-narrow 
$0_2^{+}$-resonance is treated as a bound state, whose wave function 
$\Psi^{(2)}$ is taken as a scattering solution at the resonance energy $E_2$ 
normalized on the interval $\rho \le \rho_t $, where $\rho_t \approx 45$~fm is 
the typical extention of the potential barrier. 


The solution of the five-channel system of HREs provides the desired accuracy 
for $\Gamma$, $M_{12}$, $R_{12}$, and $R_{2}$ which are presented for all 
the selected potentials in Table~\ref{tabprop}. 
Amazingly, both $\Gamma$ and $M_{12}$ coincide with 
the measurements $8.5 \pm 1.0 $~eV~\cite{Ajzenberg90} and 
$4.396 \pm 0.27$~fm$^2$~\cite{Strehl68} within the experimental errors, 
while the calculated $R_{12}$ slightly overestimates the experimental value 
$4.396 \pm 0.27$~fm~\cite{Strehl68}. 
The recent experiment~\cite{Chernykh10} of significantly improved accuracy 
gives $M_{12} = 5.47 \pm 0.09$~fm$^2$ and $R_{12} = 4.59 \pm 0.16$~fm, still 
the calculated $M_{12}$ is in agreement with these highly accurate data, 
whereas $R_{12}$ agrees even better. 
Thus, the quite natural requirements on the simple two-body and three-body 
effective potentials of the $\alpha$-cluster model turn out to 
be sufficient for a precise description of the lowest $^{12}\mathrm{C}(0^{+})$ 
states.
The reliable value of the $0^{+}_2$-state r.~m.~s. radius is found with a 
small variance about $R_2 = 3.57$~fm which exceeds $R_1$ more than $1$~fm. 
Note that the exact experimental data on $R_{2}$ are not available and 
the efforts to extract its value, e.~g., from the Fraunhofer diffraction in 
the inelastic $\alpha + ^{12}\mathrm{C}$ scattering, are affected by the model 
assumptions~\cite{Danilov09}. 

\begin{figure}[hbt]
\includegraphics[width=.8\textwidth, angle=-90]{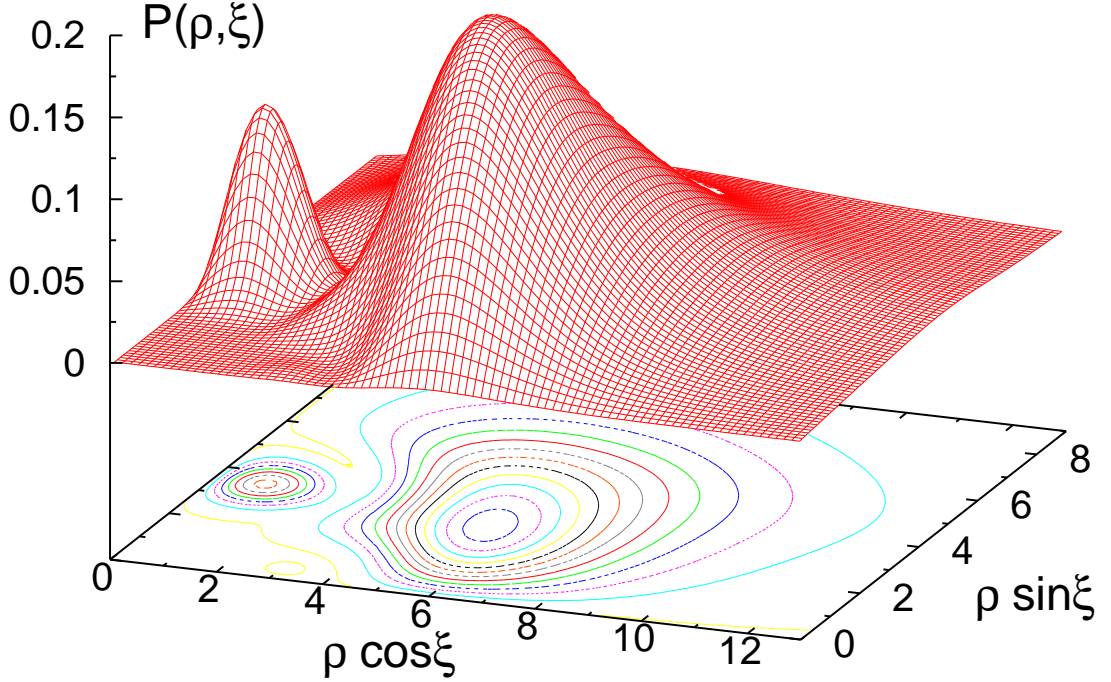}
\caption{Probability distribution of $\alpha$-particles $P(\rho, \xi )$ for 
the $^{12}\mathrm{C}(0^{+}_2)$ state. 
\label{fig2} }
\end{figure}

A stark success in the calculation of the $\Gamma$, $M_{12}$, and $R_{12}$ 
implies that the $\alpha$-cluster model provides a good description of 
the $0^{+}_2$ state's structure. 
The structural properties become more conspicuous, if represented in 
the variables $\xi $ and $\varphi_i $ defined by the expressions $\sin\xi = 
\sin\alpha_i \sin\theta_i $ and $\cos\xi \cos\varphi_i = \cos\alpha_i $. 
As the inertia momenta $I_1 = \frac{1}{2} m_\alpha \rho^2 \cos^2(\xi /2)$, 
$I_2 = \frac{1}{2} m_\alpha \rho^2 \sin^2(\xi /2)$, and 
$I_3 = \frac{1}{2} m_\alpha \rho^2$ are related to $\rho$ and $\xi $, 
the probability distribution of $\alpha$-particles $P(\rho, \xi ) = 
\frac{1}{32}\rho^5 \sin2\xi \int d \varphi |\Psi(\rho, \xi, \varphi )|^2$ is 
plotted in Fig.~\ref{fig2} to represent the features of the $0^{+}_2$ state. 
The sufficiently wide main peak corresponds to different configurations of 
large size, mostly satisfying the condition $I_2 \gg I_1$, in particular, 
$P(\rho, \xi )$ takes the maximum for the ratio $I_2/I_1 \approx 18$ and 
$\rho = 6.3$~fm. 
In other words, the elongated chain configuration is the most probable shape 
of the $0^{+}_2$ state. 
The peak extends to a ridge in the direction of increasing $I_2/I_1$ which 
clearly represents the $\alpha + \, ^8\mathrm{Be}$ decay mode. 
The minor peak is quite similar to the $0^+_1$-state probability distribution 
whose form is consistent with the equilateral triangle configuration. 
The two peaks are separated by a ravine pointing out at the position of 
the nodal surface which separates the regions of different sign of 
the wave-function. 
This means that three $\alpha$-particles in the $0^{+}_2$ state oscillate 
between the ground-state-like configuration and the linear chain 
configuration which prevails in the total probability, whereas the weight of 
the minor-peak region is about 7-9\%. 
A detailed and lengthy discusssion of the $0^{+}_2$ state structure is beyond 
the scope of this Letter and will be given elsewhere. 


In summary, it is shown that the $\alpha$-cluster model, in spite of its 
relative simplicity, provides a consistent description of 
the lowest $^{12}\mathrm{C}(0^{+})$ states. 
Even for the simple effective potentials describing the $\alpha$-$\alpha$ 
scattering data, the energy and width of $^8\mathrm{Be}$, the energies of 
$^{12}\mathrm{C}(0^{+}_{1, 2})$ states, and the $0^+_1$ r.~m.~s. radius, 
the calculated $0^{+}_2$ state width $\Gamma$, the monopole transition matrix 
element $M_{12}$, and the transition radius $R_{12}$ are in excellent 
agreement with the experimental data. 
The wave function of the $0^{+}_2$ state corresponds to the system of 
three $\alpha$-particles which oscillate between the equilateral triangle 
and the linear chain configurations, the latter contributing to the total 
probability about 90\%. 
The three-body calculation directly approves the sequential decay mechanism 
(${^{12}\mathrm{C}} \to \alpha + \, ^8\mathrm{Be} \to 3\alpha$) of 
the $0^{+}_2$ state in agreement with the experiment~\cite{Freer94}. 

The present results indicate that the theoretical calculations within 
the framework of the $\alpha$-cluster model are accurate enough to keep pace 
with the experimental data. 
There is enough room for further refinement of the $\alpha$-cluster model 
which could be used to describe both the improved experimental data and those 
reactions which depend sensitively on the fine details of the wave function. 
Important examples could be the $\alpha$-$\alpha$ bremsstrahlung and the 
($\alpha$, $\alpha$) reactions. 
Furthermore, the present approach is promising to study the triple-$\alpha$ 
reaction ($3\alpha \to {^{12}\mathrm{C}}$) at low energy below the three-body 
resonance that provides an opportunity for unified treatment of the crossover 
from the resonant to the non-resonant mechanism of the reaction. 

\bibliography{3alpha}

\end{document}